\documentclass[pra,twocolumn,showpacs,superscriptaddress,amsfonts,amsmath,,amssymb,floatfix]{revtex4}
\usepackage{graphicx}
\usepackage{psfrag}
\usepackage{epsfig}
\usepackage{pstricks}
\newcommand{\ket}[1]{| #1 \rangle}
\newcommand{\bra}[1]{\langle #1 |}
\newcommand{\braket}[2]{\langle #1 | #2 \rangle}
\newcommand{\ketbra}[2]{| #1 \rangle \langle #2 |}
\newcommand{\ketup}{| \!\! \uparrow \rangle}
\newcommand{\braup}{\langle \uparrow \!\! |}
\newcommand{\ketdown}{|\!\! \downarrow \rangle}
\newcommand{\bradown}{\langle\downarrow \!\! |}
\newcommand{\ketupup}{| \!\! \uparrow \uparrow \rangle}

\newcommand{\ketdowndown}{|\!\! \downarrow \downarrow \rangle}

\newcommand{\ketupdown}{| \!\! \uparrow \downarrow \rangle}

\newcommand{\ketdownup}{| \!\! \downarrow \uparrow \rangle}

\newcommand{\ketbradownup}{| \!\! \downarrow \rangle \langle \uparrow \!\! |}
\newcommand{\ketbraupdown}{|\!\!  \uparrow \rangle \langle \downarrow \!\! |}

\newcommand{\up}{\uparrow}
\newcommand{\down}{\downarrow}

\newcommand{\I}{{\rm{i}}}
\newcommand{\D}{{\rm{d}}}
\newcommand{\E}{e}
\newcommand{\tr}{\operatorname{tr}}
\newcommand{\trace}{\operatornamewithlimits{\tr}}

\newcommand{\nn}{\nonumber}

\def\real{{\mathbb{R}}}

\def\complex{{\mathbb{C}}}

\def\proba{{\rm I\kern -.18em P}}

\newcommand{\ie}{i.e.}

\newcommand{\vv}{{\bf{v}}}
\newcommand{\sigmav}{{\mathbf{\sigma}}}

\newcommand{\Nn}{{\cal N}}
\newcommand{\Oo}{{\cal O}}

\begin{document}

%%%%%%%%%%%%%%%%%%%%%%%%%%%%%%%%%%%%%%%%%%%%%%%%%%%%%%%%%%%%%%%%%%%%%%%%%%%%%%5
\title {Average entanglement for Markovian quantum trajectories}
\author{S. Vogelsberger}\email{vogelsy@ujf-grenoble.fr}
\affiliation{Institut Fourier, Universit\'e Joseph Fourier and CNRS, 
BP 74, 38402 Saint Martin d'H\`eres, France}
\author{D. Spehner}
%\email{spehner@ujf-grenoble.fr}
\affiliation{Institut Fourier, Universit\'e Joseph Fourier and CNRS, 
BP 74, 38402 Saint Martin d'H\`eres, France}
\affiliation{Laboratoire de Physique et Mod\'elisation des
Milieux Condens\'es, Universit\'e Joseph Fourier and CNRS, BP 166, 38042 Grenoble, France}
%\date{\today}

%%%%%%%%%%%%%%%%%%%%%%%%%%%%%%%%%%%%%%%%%%%%%%%%%%%%%%%%%%%%%%%%%%%%%%%%%%%%5
\begin{abstract}
We study the evolution of the entanglement of noninteracting qubits coupled to
reservoirs under monitoring of
the reservoirs by means of continuous measurements.
We calculate the average 
of the concurrence of the qubits wavefunction over all quantum trajectories.
For two qubits coupled to independent baths subjected to local measurements, 
this average
decays exponentially with a rate depending on the measurement scheme
only.
This contrasts with the known  disappearance  of entanglement 
after a finite time 
for the density matrix in the absence of measurements.
For two qubits coupled to a common bath, the mean concurrence can vanish
at discrete times. 
Our analysis applies to arbitrary quantum jump or quantum state diffusion 
dynamics in the Markov limit.
We discuss the best measurement schemes to protect entanglement in specific examples. 
 \end{abstract}
\maketitle
%=====================================================================================
%%%%%%%%%%%%%%%%%%%%%%%%%%%%%%%%%%%%%%%%%%%%%%%%%%%%%%%%%%%%%%%%%%%%%%%%%%%%%%
\section{Introduction}
%%%%%%%%%%%%%%%%%%%%%%%%%%%%%%%%%%%%%%%%%%%%%%%%%%%%%%%%%%%%%%%%%%%%%%%%%%%%%%%

Entanglement is a key resource in quantum information. It can
be  destroyed or sometimes created by interactions  with
a reservoir.
When the two non-interacting parts of 
a bipartite system are coupled to 
 independent baths, entanglement typically disappears  after a 
finite time~\cite{Diosi04,Dodd04,Eberly04,Almeida07}. This phenomenon,
called ``entanglement sudden death'' (ESD),  
occurs  for certain initial states only or for all entangled  initial  states, depending 
on whether the system relaxes
to a steady state belonging to the boundary of the set of 
separable states (e.g., to a separable pure  state for 
baths at zero temperature) or  to its interior 
(e.g., to a Gibbs state at positive temperature)~\cite{Terra_Cunha07}.
A quantum state lies on this boundary if it is separable and  an arbitrarily 
small perturbation makes it entangled; this is the case, for example, 
for a pure separable state.
When the two parts of the system are coupled to a 
common bath, sudden revivals of entanglement may take place after the state has
become separable~\cite{Braun02,Ficek06,Mazzola09}.

In this article we consider the loss of entanglement between two non-interacting qubits
coupled to one or two baths monitored by continuous 
measurements. Because of these measurements, the qubits remain  at all times 
in a pure state $\ket{\psi(t)}$, which evolves randomly.
To each  measurement result (or ``realization'') corresponds 
a quantum trajectory $t \in \real_+ \mapsto \ket{\psi(t)}$ 
in the Hilbert space $\complex^4$ of the qubits.    
In the Born-Markov regime, the dynamics is given by 
the quantum jump (QJ) model~\cite{Dalibard92,Carmichael} or, in the case 
of homodyne and heterodyne detections, by
the so-called quantum state diffusion (QSD) models~\cite{Carmichael,Wiseman93,Gisin92}. 
We study how  the entanglement of the qubits evolves in time 
by calculating the average $\overline{C_{\psi(t)}}$ of the Wootters concurrence of $\ket{\psi (t)}$   
over all quantum trajectories;   
$\overline{C_{\psi(t)}}$ differs in general from the concurrence 
$C_{\rho(t)}$ of the density matrix
$\rho(t)= \overline{\ketbra{\psi(t)}{\psi(t)}}$
(here and in what follows the overline denotes the mean over all quantum trajectories)~\cite{Nha04,Carvalho}.
For two qubits coupled to
{\it independent} baths, we find that
\begin{equation} \label{eq-main_result}
\overline{C_{\psi(t)}} = C_0    \,e^{-\kappa t}
\end{equation}
where $C_0 = C_{\psi(0)}$ is the initial concurrence and
$\kappa\geq 0$ depends on the measurement scheme but
not on the  initial state $\ket{\psi(0)}$. 
In particular, if  $C_0>0$ and $t_{\rm ESD} \in ]0,\infty[$ is the time
at which entanglement disappears in the density matrix (assuming that this time is finite), 
then $C_{\rho(t)}=0$ at times $t\geq t_{\rm ESD}$ whereas 
$\overline{C_{\psi(t)}}$ can only vanish asymptotically.  
The continuous measurements on the two baths  
thus protect on average the qubits from ESD.  
Of course, this does not mean that {\it all}
random wavefunctions $\ket{\psi(t)}$ remain entangled at all times.
But in some cases, such as for pure dephasing or for
infinite temperature baths, one can find measurement schemes such that 
$\kappa =0$; then, for all trajectories, 
if the qubits are maximally entangled at $t=0$
they remain maximally entangled at all times.
We show that the best measurement scheme to protect entanglement
is in general given by homodyne detection with appropriately chosen laser phases.
Related strategies 
using quantum Zeno effect~\cite{Maniscalco08},
entanglement distillation~\cite{Mundarain09}, quantum feedback~\cite{CarvalhoPRA07}, 
and encoding in qutrits~\cite{Mascarenhas10} have been proposed.
It is assumed in this work that the  measurements on the baths are performed by 
perfect detectors. The  impact of detection errors has been studied in~\cite{Mascarenhas10}.

When  the qubits are coupled  to a common bath, we find that
$\overline{C_{\psi(t)}}$ has a more complex time behavior than in (\ref{eq-main_result}). It
may vanish at finite discrete times, and, for some initial states, 
be equal to $C_{\rho(t)}$.
It is worthwhile to stress that 
the formula (\ref{eq-main_result}) is valid 
provided not only each qubit is coupled to its own bath,
but also the baths are monitored  independently from each other by the measurements.
This means that the measurements are performed {\it locally} on each bath. 
Instead of looking for the measurement scheme maximizing 
the average concurrence  $\overline{C_{\psi(t)}}$ of the two qubits 
in order to obtain the best entanglement protection,
it is also of interest to find a way to perform the measurements such that 
$\overline{C_{\psi(t)}}$ is minimum and  
coincides with the  concurrence $C_{\rho(t)}$ of the density matrix. This problem 
has been studied  numerically in Ref.~\cite{Carvalho} and  analytically 
in~\cite{Viviescas-arXiv} for specific models of
couplings with the two baths.
Our result (\ref{eq-main_result}) implies that for any Markovian dynamics, 
if the two qubits are initially entangled and   ESD occurs for
 the density matrix $\rho(t)$, a scheme 
with the aforementioned property must necessarily
involve measurements of non-local (joint) observables 
of the two baths.
In the models studied in~\cite{Carvalho,Viviescas-arXiv}, non-local measurements
are indeed used in order to obtain an optimal scheme satisfying 
$\overline{C_{\psi(t)}}=C_{\rho(t)}$.

The paper is organized as follows. We briefly recall in Sec.\ \ref{sec-entanglement_measure} 
 the definition of the
concurrence of pure and mixed states and review the quantum jump unraveling 
of a Lindblad  equation for the density matrix in Sec.\ \ref{sec-QJM}.
We treat the  simple and illustrative case
of two two-level atoms coupled to independent baths at zero 
temperature in Sec.\ \ref{sec-photon_counting}, before showing  
formula (\ref{eq-main_result}) in Sec.\ \ref{sec-proof_main_result} 
for a general quantum jump dynamics.
The QSD unravelings are considered in Sec.\ \ref{sec-QSD};
we obtain  
the average concurrence for such unravelings as limits of the concurrence
for QJ dynamics (corresponding to homodyne and heterodyne detections 
with intense laser fields).
Section \ref{sec-common_bath} is devoted to the evolution of the entanglement
of two qubits coupled to a common 
bath at zero temperature. The main conclusions of the work are given in 
Sec.\ \ref{sec-conclusion}.

%%%%%%%%%%%%%%%%%%%%%%%%%%%%%%%%%%%%%%%%%%%%%%%%%%%%%%%%%%%%%%%%%%%%%%
\section{Entanglement measures for quantum trajectories}
\label{sec-entanglement_measure}
%%%%%%%%%%%%%%%%%%%%%%%%%%%%%%%%%%%%%%%%%%%%%%%%%%%%%%%%%%%%%%%%%%%%%%

The entanglement of formation of a
bipartite quantum system ${\cal S}$ in a pure state $\ket{\psi}$
 is defined by means of the von Neumann entropy
$E_\psi = - \tr (\rho_A \ln \rho_A)=-\tr (\rho_B \ln \rho_B)$ 
of  the density matrices  $\rho_A = \tr_{B}( \ketbra{\psi}{\psi})$  
and $\rho_B = \tr_{A}( \ketbra{\psi}{\psi})$ of the two subsystems 
$A$ and $B$ composing ${\cal S}$~\cite{Bennett96}.
If ${\cal S}$ is in a mixed state,
$E_\rho$ is  the infimum of
$\sum p_k E_{\psi_k}$ over all convex decompositions 
$\rho = \sum_k p_k \ketbra{\psi_k}{\psi_k}$ of its density  matrix
(with $p_k \geq 0$ and $\| \psi_k\|=1$). 
When $A$ and $B$  have  two-dimensional 
Hilbert spaces, $E_\rho = f ( C_\rho)$  is related to the concurrence~\cite{Wootters98}
$C_\rho$ by a convex  increasing function $f: [0,1]\rightarrow [0,\ln (2)]$;
$\rho$ is separable if and only if $C_\rho=0$, \ie, $E_\rho=0$.
For a pure state~\cite{Wootters98}, 
\begin{equation} \label{eq-def-concurrence}
C_\psi = | \langle \sigma_{y} \otimes \sigma_{y} T \rangle_{\psi} |
\end{equation}
where $\sigma_y = \I (\ketbradownup - \ketbraupdown )$ is the 
$y$-Pauli matrix, $T: \ket{\psi}
 = \sum_{s,s'} c_{s s'} \ket{s,s'} \mapsto 
 \sum_{s,s'} c_{s s'}^\ast \ket{s,s'}$
the  anti-unitary operator of complex conjugation
in the canonical 
basis $\{\ket{s,s'} = \ket{s} \otimes \ket{s'} ; s,s' =\uparrow,\downarrow\}$ of 
$\complex^2 \otimes \complex^2$, and 
$\langle \cdot \rangle_{\psi} = \bra{\psi} \cdot \ket{\psi}$ the quantum expectation 
in state $\ket{\psi}$.

For quantum trajectories, 
one has always  $\overline{E_{\psi (t)}} \geq E_{\rho (t)}$, this inequality being strict
excepted if  the decomposition 
\begin{equation} \label{eq-def_rho}
\rho (t) = \overline{\ketbra{\psi(t)}{\psi (t)}}
= \int \D p [\psi] \,\ketbra{\psi(t)}{\psi (t)}
\end{equation}
realizes the infimum defining $E_{\rho (t)}$.
Thanks to the convexity of $f$,
$\overline{E_{\psi(t)}} \geq f(\overline{C_{\psi(t)}} )$.
Thus equation~(\ref{eq-main_result}) shows that for independent baths and if $C_0>0$,  
$\overline{E_{\psi(t)}}\geq f( C_0 \E^{-\kappa t})>0$ 
whatever the measurement scheme. 

It is legitimate to ask which entanglement measure
should be averaged since, for example, 
$\overline{E_{\psi(t)}}=E_0$ could be constant
and $\overline{C_{\psi(t)}}$ time-decreasing
if  $E_0 \not= 0, \ln 2$. 
The concurrence is a natural candidate as it corresponds for  pure states to the supremum
 over all self-adjoint  local observables $J_A$ and $J_B$ with norms 
less than one  
of the modulus of the correlation between $J_A$ and $J_B$,
\begin{eqnarray}
\nonumber
C_{\psi (t)}
& = &  
\sup_{\|J_A\|, \|J_B\| \leq 1}
\bigl| \langle J_A \otimes J_B \rangle_{\psi(t)} 
\\
& & - \langle J_A \otimes 1_B\rangle_{\psi (t)} \langle 1_A \otimes J_B \rangle_{\psi (t)} \bigr|\,.
\end{eqnarray}
 Moreover,  
$\overline{C_{\psi (t)}}$ is easy to calculate in the Markov regime and gives a 
lower bound on $\overline{E_{\psi (t)}}$.

%%%%%%%%%%%%%%%%%%%%%%%%%%%%%%%%%%%%%%%%%%%%%%%%%%%%%%%%%%%%%%%%%%%
\section{Quantum jump model} 
\label{sec-QJM}
%%%%%%%%%%%%%%%%%%%%%%%%%%%%%%%%%%%%%%%%%%%%%%%%%%%%%%%%%%%%%%%%%%%%%

Let us briefly recall the QJ dynamics~\cite{Dalibard92,Breuer,Knight98}.
As a result of a measurement
on a particle (e.g. a photon) of the bath scattered by the qubits,
the qubits wavefunction suffers a quantum jump
\begin{equation} \label{eq-jump_dyn}
\ket{\psi(t)} 
 \longrightarrow 
  \ket{\psi_{\text{jump}}^{(m,i)}} = \frac{J_m^i \ket{\psi(t)}}{\| J_m^i \ket{\psi(t)}\| }
\end{equation}
where the jump operator $J_m^i$ is related to the
particle-qubits coupling and the indices $m,i$ label all 
possible measurement results save for the most likely one, which we call a 
``no detection''.
In the weak coupling limit, 
the probability that a measurement in the small time interval $[t, t+\D t]$
gives the result $(m,i)$ is  very small and equal to
$\D p_m^i (t) = \gamma_m^i \| J_m^i \ket{\psi(t)}\|^2 \D t$. The jump rate 
$\gamma_m^i$ does not depend
on $\ket{\psi(t)}$ and is proportional to the square of the particle-qubit coupling constant. 
In the no-detection case the wavefunction of the qubits evolves as  
\begin{equation}  \label{eq-no_jump}
\ket{\psi(t+\D t)} 
\! = \! 
  \frac{e^{-\I H_{\rm{eff}} \D t } \ket{\psi(t)}}
   {\| e^{-\I H_{\rm{eff}}\D t } \ket{\psi(t)}\|}
,
H_{\rm{eff}}\!=\! H_0-\!\frac{\I}{2} \sum_{m,i} \gamma_m^i J_m^{i \dagger} J_m^i 
\end{equation}
where $H_0$ is the Hamiltonian of the qubits. 
The probability that no jump occurs  
in the time interval $[t_0,t]$ is 
$p_{\text{nj}} (t_0,t)=\| e^{-\I H_{\rm{eff}} (t-t_0)} \ket{\psi (t_0)}\|^2 $.
(This is proven as follows: 
as $p_{\text{nj}} (t_0,t) -p_{\text{nj}} (t_0,t + \D t)  = \sum_{m,i} \D p_m^i (t) 
p_{\text{nj}} (t_0,t)$, one has 
$\partial \ln p_{\text{nj}}(t_0,t) /\partial t
 =$
$ -  
\sum_{m,i}
\gamma_{m}^{i} \| J_m^i \ket{\psi(t)} \|^2
 = 
(\partial /\partial t) \ln \| e^{-\I H_{\rm{eff}} (t-t_0)} \ket{\psi (t_0)}\|^2$
by (\ref{eq-no_jump}).)
It is not difficult to 
show~\cite{Dalibard92} that
the  density matrix 
$\rho(t) =\overline{\ketbra{\psi(t)}{\psi (t)}}$ satisfies the Lindblad 
equation
\begin{equation} \label{eq-Lindblad}
\frac{\D \rho}{\D t}  =  
- \I [ H_0, \rho ] 
 +  \sum_{m,i} \gamma_m^i \Bigl( J_m^i \rho J_m^{i \dagger} 
 - \frac{1}{2} \bigl\{ J_m^{i \dagger} J_m^i , \rho  \bigr\} \Bigr)
\end{equation}
where $\{ \cdot , \cdot \}$ denotes the anti-commutator.
It is known that
many distinct QJ dynamics unravel the same 
master equation (\ref{eq-Lindblad})~\cite{Breuer}. 
For two qubits coupled to independent reservoirs $R_A$ and $R_B$, the 
jump operators are {\it local}, \ie, they have the form 
\begin{equation} \label{eq-jump_op}
J_m^A \otimes 1_B
\quad , \quad 
1_A \otimes J_{m}^B
\end{equation}
depending on whether the measurements are performed on $R_A$ or $R_B$. Here 
$J_m^i$ are $2 \times 2$ matrices. 

The aforementioned absence of ESD for the mean concurrence 
of two qubits coupled to independent baths can be traced back to the existence of trajectories
for which $\ket{\psi(t)}$  remains entangled at all times.
Actually, for a trajectory 
without jump, 
$\ket{\psi_{\rm{nj}} (t)} \propto \E^{-\I t H_{\rm eff}}\ket{\psi (0)}$, 
see (\ref{eq-no_jump}). 
By (\ref{eq-jump_op}) and since the qubits do not interact with each other, 
$e^{-\I t H_{\rm eff}}$ is the tensor product of two local operators acting on each qubit. 
If $\ket{\psi_{\rm{nj}}(t)}$ would be 
separable at a given
 time $t$ then, by reversing the dynamics (\ie, by applying $\E^{\I t H_{\rm eff}}$ to 
$\ket{\psi_{\rm{nj}}(t)}$) one would deduce that $\ket{\psi (0)}$ is separable. 
Hence $C_{\psi_{\rm{nj}}(t)}>0$ if $C_0>0$. 
But the no-detection probability between times $0$ and $t$ is nonzero and
thus $\overline{C_{\psi(t)}}>0$ at all times. 
Note that this argument does not apply  if non-local observables of the two baths 
are measured or if the two qubits are coupled to a common bath,
since then the jump operators are non-local. 

%%%%%%%%%%%%%%%%%%%%%%%%%%%%%%%%%%%%%%%%%%%%%%%%%%%%%%%%%
\section{Photon counting}
\label{sec-photon_counting}
%%%%%%%%%%%%%%%%%%%%%%%%%%%%%%%%%%%%%%%%%%%%%%%%%%%%%%%%%%%

Let us illustrate the random dynamics described previously  on a simple and 
experimentally relevant example~\cite{Haroche}.
Each qubit is a two-level atom 
coupled resonantly to the electromagnetic 
field initially in the vacuum  (zero-temperature photon bath). 
The two atoms are far from each other and thus  
interact with independent field modes.
Two perfect photon counters $D_i$ make a click when a photon is emitted by qubit $i$ 
($i=A,B$),
whatever the direction of the emitted photon. Doing  the rotating wave approximation, 
the jump operators
are $J^i_{-} = \sigma_{-}^i= \ketbradownup$. For simplicity we take $H_0=0$.
 By~(\ref{eq-no_jump}), if no photon is detected in the time interval $[0,t]$ the qubits state 
is 
\begin{equation}
\ket{\psi (t)} 
 =  \Nn (t)^{-1}\sum_{s,s'=\uparrow ,\downarrow} c_{s s'}  \,\E^{-\gamma_{s s'} t/2}\,\ket{s,s'}
\end{equation}
with  $\gamma_{\uparrow \uparrow} = \gamma_A + \gamma_B$,
$\gamma_{\uparrow \downarrow} = \gamma_A$, $\gamma_{\downarrow \uparrow} = \gamma_B$, 
 $\gamma_{\downarrow \downarrow} = 0$ 
($\gamma_i$ being  the jump rate for detector $D_i$),
$c_{s s'} = \braket{s,s'}{\psi(0)}$, and
$\Nn (t)^2 = 
\sum_{s,s'} |c_{s s'} |^2 \E^{-\gamma_{s s'} t}$.
The concurrence (\ref{eq-def-concurrence}) of $\ket{\psi (t)}$  is
$C(t) = C_0 \,\Nn (t)^{-2} e^{-(\gamma_A+\gamma_B) t/2}$
with $C_0 = 2 | c_{\uparrow \uparrow} c_{\downarrow \downarrow} 
-  c_{\uparrow \downarrow} c_{\downarrow \uparrow}|$.  
If a photon is detected at time $t_j$ by, say, the photon counter $D_A$,
the qubits are just after
the jump (\ref{eq-jump_dyn}) in the separable state 
$\ket{\psi(t_j +)} 
 \propto$  
$\ketdown \otimes ( c_{\uparrow \uparrow} e^{-\gamma_{\uparrow \uparrow}  t_j/2} {\ketup} 
+ c_{\uparrow \downarrow} e^{-\gamma_{\uparrow \downarrow} t_j/2} \ketdown )$. 
Since neither a jump  nor the inter-jump dynamics
can create entanglement (the jump operators~(\ref{eq-jump_op}) being local),
$\ket{\psi(t)}$ remains 
separable at all times $t \geq t_j$, even if more photons are subsequently 
detected.     
Thus $C(t) =0$ if at least one photon is detected in the time interval $[0,t]$.
Averaging over all realizations of the quantum trajectories and using
the probability  $p_{\text{nj}} (0,t)=\Nn(t)^2$ that no photon is detected
% by  $D_A$ and $D_B$ 
in $[0,t]$,
one finds $\overline{C(t)} =C_0 e^{-(\gamma_A + \gamma_B) t/2}$.
   
\begin{figure}
\includegraphics*[width=1\columnwidth]{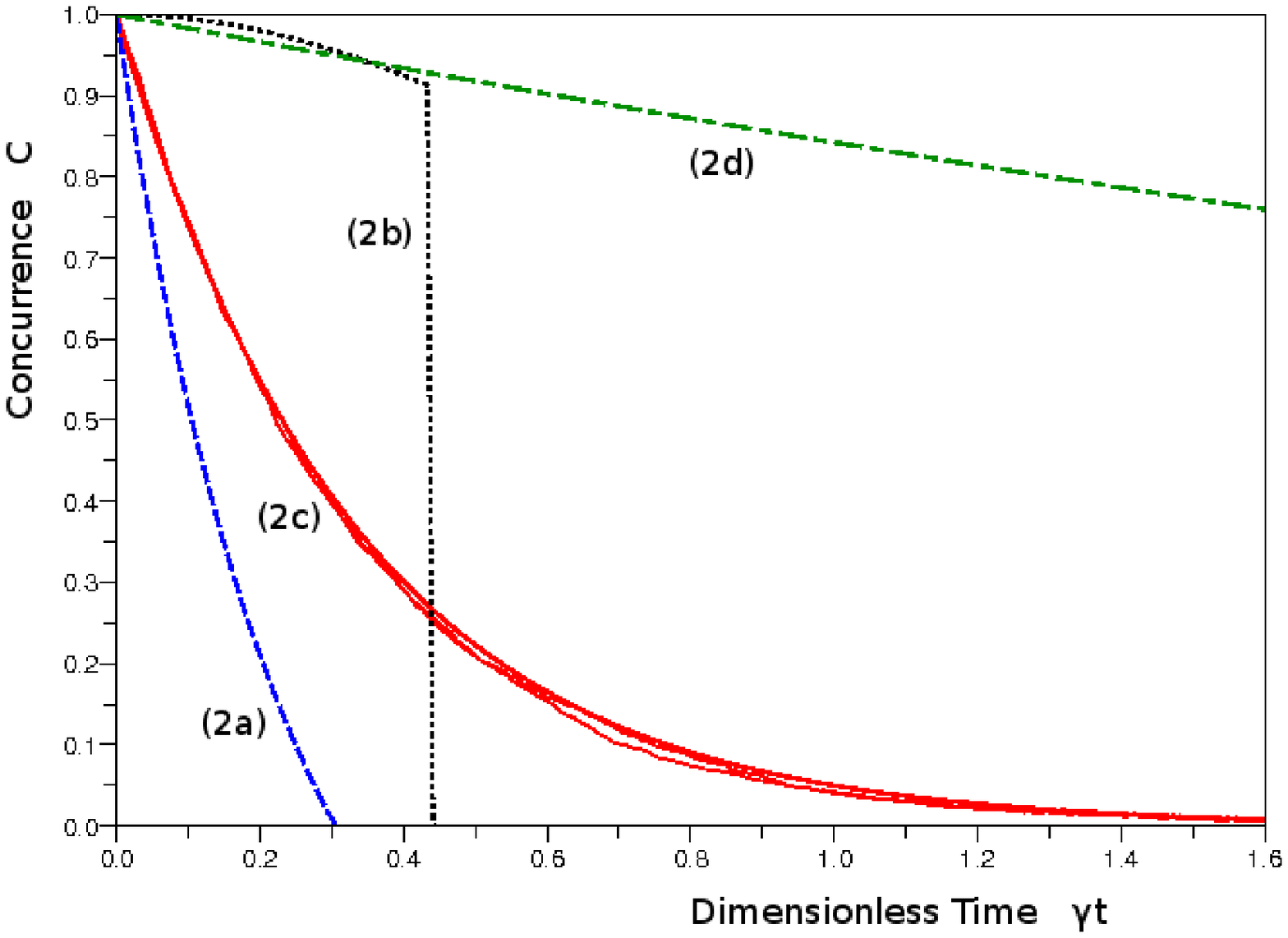}
\caption{\label{fig-concurrence_2_baths_T>0}
(Color online) 
Concurrences of two qubits
coupled to independent baths 
at positive temperature  as a function of $\gamma t$
for $\gamma_{+}^i  = \gamma_{-}^i/2=\gamma$ and
$\ket{\psi(0)}=(\ketupup -\I  \ketdowndown )/\sqrt{2}$:
(2a)~$C_{\rho(t)}$ for the density matrix (blue dashed line);  
(2b)~$C_{\psi (t)}$  for a single trajectory 
(black dotted line);
(2c)~$\overline{C_{\psi (t)}}$ averaged over 1500 trajectories  
and from Eq.~(\ref{eq-main_result}) (red solid lines);
(2d) $\overline{C_{\psi (t)}}$ for the best measurement
scheme (see text).
}
\end{figure}

This argument is easily extended to baths 
at positive temperatures by adding  
two jump operators $J_{+}^i=\sigma_{+}^i$ 
with rates 
$\gamma_+^i \leq  \gamma_{-}^i$. 
The mean concurrence 
is then 
$
\overline{C(t)} =C_0 e^{-(\gamma_+^A + \gamma_-^A + \gamma_+^B + \gamma_-^B ) t/2}
$.
It is compared in  Fig.\ \ref{fig-concurrence_2_baths_T>0}
with the concurrence of the density matrix obtained by solving 
the master equation 
(\ref{eq-Lindblad}), which shows ESD for all initial states.

%%%%%%%%%%%%%%FIGURE 1 %%%%%%%%%%%%%%%%%%%%%%%%%%%%%%%%%%%%%%%%%%%%%
%%%%%%%%%%%%%%%%%%%%%%%%%%%%%%%%%%%%%%%%%%%%%%%

%%%%%%%%%%%%%%%%%%%%%%%%%%%%%%%%%%%%%%%%%%%%%%%%%%%%%%%%%%%%%%
\section{General quantum jump dynamics}
\label{sec-proof_main_result}
%%%%%%%%%%%%%%%%%%%%%%%%%%%%%%%%%%%%%%%%%%%%%%%%%%%%%%%%%%%%%%

We now consider a general QJ dynamics with  jump operators given 
by (\ref{eq-jump_op}). The Hamiltonian of the qubits has the form
$H_0 = H_{A} \otimes 1_B + 1_A\otimes H_{B}$.
Let 
$K= K_A \otimes 1_B + 1_A \otimes K_B$ with
\begin{equation}
K_i=\frac{1}{2}  \sum_m \gamma_{m}^{i}\, J_m^{i \dagger} J_m^i\,,
\end{equation}
$\gamma_{m}^{i}$ being the jump rates for the detector $D_i$ ($i=A,B$). 
We first assume that no jump occurs between $t$ and $t+\D t$.
By expanding the exponential in (\ref{eq-no_jump}), one gets
\begin{eqnarray} \label{eq-C_nj}
& &  C (t+\D t) 
 = 
p_{\text{nj}} (t,t+\D t)^{-1} 
\bigl| \bigl\langle \sigma_{y} \otimes \sigma_{y} T \bigr\rangle_{\psi (t)} 
\\
& & 
\nn
+\I \D t \bigl\langle H_{\rm{eff}}^\dagger \sigma_{y} \otimes \sigma_{y} T 
+ \sigma_{y} \otimes \sigma_{y} T H_{\rm{eff}} \bigr\rangle_{\psi (t)} + \Oo (\D t)^2\bigr| 
\end{eqnarray}
where  $p_{\text{nj}} (t,t+\D t)= 
\langle 1 - 2 K  \D t +\Oo (\D t)^2 \rangle_{\psi (t)}$ 
is the probability that no jump occurs between $t$ and $t + \D t$. 
Now, for any local operator $O_i$ acting on qubit $i$, one has
\begin{equation}
\bigl\langle O_i \sigma_{y} \otimes \sigma_{y} T \bigr\rangle_{\psi (t)}
\!\!=\! \bigl\langle \sigma_{y} \otimes \sigma_{y} T O_i^\dagger \bigr\rangle_{\psi (t)}
\!\!=\! \frac{ {\cal{C}} (t)}{2}{\tr}_{\complex^2} (O_i)
\end{equation}
 with
\begin{equation} \label{eq-D(t)}
{\cal{C}} (t) = \langle \sigma_{y} \otimes \sigma_{y} T \rangle_{\psi (t)}
\!= \!2 \bigl(  
c_{\uparrow \downarrow}^\ast (t) c_{\downarrow \uparrow}^\ast (t) 
- c_{\uparrow \uparrow}^\ast(t) c_{\downarrow \downarrow}^\ast (t) 
\bigr)
\end{equation}
and $ c_{s s'} (t) = \braket{s,s'}{\psi (t)}$. Since 
$C(t) = | {\cal{C}} (t)|$ and $H_{\rm{eff}}=\sum_i (H_i - \I K_i)$, this gives
\begin{equation} \label{eq-conc_increment_no_jump}
 C (t+\D t) \,p_{\text{nj}} (t,t+\D t) 
= C (t) \Bigl( 1 - \tr_{\complex^4} (K)  \frac{\D t}{2}   + \Oo (\D t^2) \Bigr)
\,.
\end{equation}
If detector $D_i$  gives the result $m$ in the time interval $[t,t + \D t]$,
the concurrence is by virtue of~(\ref{eq-jump_dyn})
\begin{equation} \label{eq-conc_increment_jump}
C_{\text{jump}}^{(m,i)} (t+\D t) 
=
\frac{\gamma_{m}^{i} \,\D t}{\D p_{m}^{i} (t)} C(t) 
\bigl| {\det}_{\complex^2} (J_m^i) \bigr|
\,,
\end{equation}
where we have used the identity
\begin{equation}
\langle O_i^\dagger \sigma_y \otimes \sigma_y T O_i \rangle_{\psi (t)} 
= {\cal{C}}(t) {\det}_{\complex^2}(O_i^\dagger )
\end{equation}
valid for any local operator $O_i$ acting on qubit $i$.
Collecting the previous formulas and using the Markov property of the jump process, 
one gets
$\overline{C(t+\D t)} = \overline{C(t)} ( 1 - \kappa_{\rm{QJ}} \,\D t + \Oo (\D t^2))$
with
\begin{equation} \label{eq-kappa}
\kappa_{\rm{QJ}} 
  =   
  \frac{1}{2} \tr_{\complex^4} ( K ) - 
   \sum_{m,i} \gamma_{m}^{i}
    | {\det}_{\complex^2} (J_m^i) |  \,.
\end{equation}
Letting $\D t$ go to zero, one obtains 
$\D \overline{C(t)}/\D t = - \kappa_{\rm{QJ}} \,\overline{C(t)}$. The solution
(\ref{eq-main_result}) of this differential equation has the exponential decay 
claimed previously. To show that $\kappa_{\rm{QJ}} \geq 0$, let
$2\theta_m^i$ be the argument of $\det_{\complex^2} (J_m^i)$. We write
$\kappa_{\rm{QJ}}= \sum_{m,i} \gamma_m^i 
[ {\tr}_{\complex^2} (J_m^{i \dagger} J_m^i )  - 2 \Re \{ \E^{-2 \I \theta_m^i}\,
{\det}_{\complex^2} (J_m^i)\}]/2$ as
\begin{equation} \label{eq-kappa_m}
\kappa_{\rm{QJ}} 
 =  
 \sum_{m,i} \frac{\gamma_m^i}{2} 
 \Bigl( 
  \bigl| 
   \braup \tilde{J}^i_{m} \ketup -  \bradown \tilde{J}^{i \dagger}_{m} \ketdown
 \bigr|^2
+ 
\bigl| 
 \braup 2 \Re \tilde{J}^i_{m} \ketdown 
\bigr|^2
\Bigr)
\end{equation}
with $\tilde{J}^i_{m}= \E^{-\I \theta_m^i} J_m^i$ and 
$2 \Re\tilde{J}^i_{m}= \tilde{J}^i_{m}+\tilde{J}^{i \dagger}_{m}$. 
Thus $\kappa_{\rm{QJ}}$ is non-negative.

\begin{figure}
\includegraphics*[width=1\columnwidth]{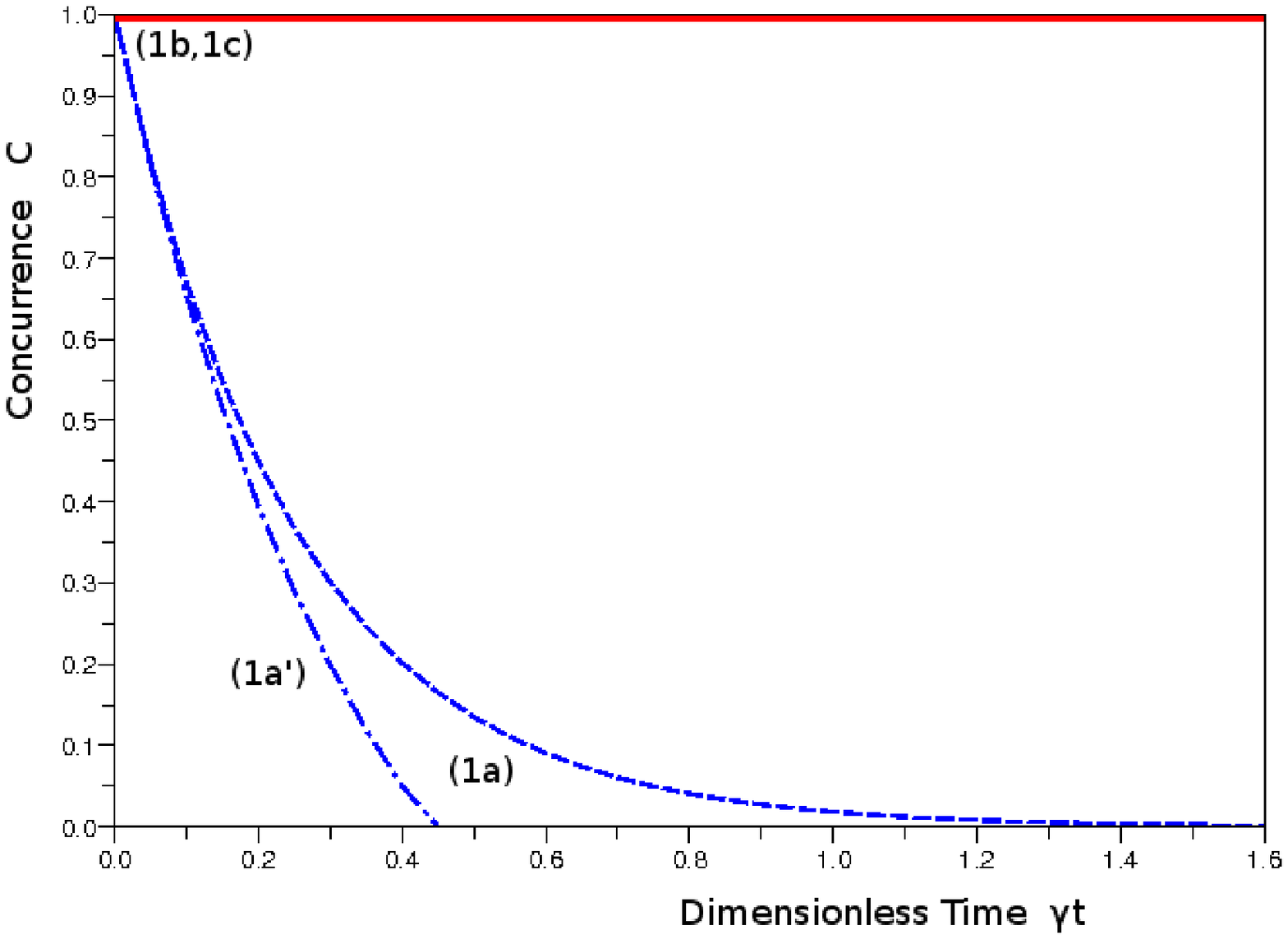}
\caption{\label{fig-concurrence_2_baths_deph}
(Color online) 
Same as in Fig.\ref{fig-concurrence_2_baths_T>0}
for
pure dephasing and the
initial state $\ket{\psi(0)}=\frac{1}{\sqrt{2}}(\ketupup + e^{-\I \varphi} \ketdowndown )$:
(1a)~$C_{\rho(t)}$ for
$\varphi=\frac{\pi}{2}$ (blue dashed line); (1a')~$C_{\rho(t)}$ for $\varphi=0$ 
(blue line showing ESD);
(1b,1c)~$C_{\psi (t)}=\overline{C_{\psi (t)}}$ (red solid line).
}
\end{figure}

Note that  $\kappa_{\rm{QJ}} =0$ if all matrices $J_m^i$ are 
 self-adjoint and traceless (then $\theta_m^i=\pi/2$ and $\Re \tilde{J}^i_{m}=0$).
We  show in Fig.\ \ref{fig-concurrence_2_baths_deph}
the concurrence of 
the density matrix given by solving (\ref{eq-Lindblad}) for a pure dephasing
with $J^i = \E^{\I \pi/4} \sigma_{-}^i + \E^{-\I \pi/4} \sigma_{+}^i$. 
One has ESD for all initial states save for 
$\ket{\psi (0)} = (\ketupup \pm \I \ketdowndown )/\sqrt{2}$.
Since $\kappa_{\rm{QJ}}=0$, (\ref{eq-main_result}) implies 
$\overline{C (t)} = C_0$.
If the two qubits are maximally entangled at $t=0$, 
then $C_{\psi (t)}=\overline{C (t)} = C_0 = 1$ for {\it all} quantum trajectories at any 
time $t \geq 0$.
Therefore, for pure dephasing  one can
protect perfectly the qubits by measuring continuously and locally the two 
independent baths.  

We can now give the optimal measurement scheme to protect the entanglement of  
two qubits coupled to independent baths at positive temperatures.
Let us replace the photon-counting jump operators $J_\pm^i = \sigma_\pm^i$ by
$J_\mu^i = \sum_{m=\pm} (\gamma^i_{m}/\gamma_\mu^i)^{\frac{1}{2}} u_{\mu m}^i \sigma_m^i$
where $U_i=(u_{\mu m}^i)_{\mu=1,\cdots, N}^{m=\pm}$ are unitary $2 \times N$ matrices.
This corresponds to a rotation of the  measurement basis and gives another 
unraveling of the master equation (\ref{eq-Lindblad}).
Let us stress that the new jump operators $J_\mu^i$ still act locally on each qubit.
By (\ref{eq-kappa}), the new rate is  
$\kappa 
= \sum_{\mu,i} ( \sqrt{\gamma_{-}^i} | u_{\mu -}^i | - \sqrt{\gamma_+^i} | u_{\mu+}^i | )^2/2$.
By using 
$\sum_\mu |u_{\mu\,\pm}^i |^2 = 1$ and optimizing over all 
unitaries $U_i$, 
one finds 
that the smallest disentanglement rate arises when, for example, 
$u_{1\,\pm}^i=\pm u_{2\,\pm}^i=1/\sqrt{2}$ ($N=2$) and is given by
\begin{equation} \label{eq-optimal_kappa_T>0}
\kappa_{\rm{QJ}}^{\rm{opt}} =
\frac{1}{2} \sum_{i=A,B} \Bigl( \sqrt{\gamma_{-}^i} - \sqrt{\gamma_{+}^i} \Bigr)^2\,.
\end{equation}
Note that 
${\kappa}_{\rm{QJ}}^{\rm{opt}}={\kappa}_{\rm{QJ}}$ at zero temperature
and ${\kappa}_{\rm{QJ}}^{\rm{opt}}=0$ (perfect protection) at infinite temperature.
The decay of $\overline{C(t)}$ for this optimal measurement is shown in  
Fig.\ \ref{fig-concurrence_2_baths_T>0} (green dashed-dotted line). 

%%%%%%%%%%%%%%FIGURE 2 %%%%%%%%%%%%%%%%%%%%%%%%%%%%%%%%%%%%%%%%%%%%%
%%%%%%%%%%%%%%%%%%%%%%%%%%%%%%%%%%%%%%%%%%%%%%%%%%%%%%%%

%%%%%%%%%%%%%%%%%%%%%%%%%%%%%%%%%%%%%%%%%%%%%%%%%%%%%%%%%%%%%%%%
\section{Homodyne and heterodyne detection}
\label{sec-QSD}
%%%%%%%%%%%%%%%%%%%%%%%%%%%%%%%%%%%%%%%%%%%%%%%%%%%%%%%%%%%%%%%%

Let us come back to our example of two atoms coupled to the electromagnetic 
field initially in the vacuum.
If homodyne photo-detection is used instead of photon counting,
the jump operators become
$J^{i}_{\pm \alpha}= \sigma_{-}^i \pm  \alpha_i$, 
 $\alpha_i$ being the amplitude of a classical laser field 
(there are now four jump operators
since each homodyne detector involves two photon counters)~\cite{Wiseman93}.
Assuming that the two photon beams emitted by the atoms  
are combined with the two laser fields via 50\% beam splitters,
the jump rates associated with $J^{i}_{\pm \alpha}$ are equal, 
$\gamma^{i}_{\pm \alpha} =\gamma_i/2$.
Thanks to (\ref{eq-kappa}), one easily finds that
the disentanglement rate  for the new QJ dynamics,
$\kappa_{\rm{QJ}} (\alpha)=(\gamma_A + \gamma_B)/2$, is the same as for photon counting. 

In contrast, $\kappa_{\rm{QJ}} (\alpha)$ depends on the laser
amplitudes for pure dephasing (jump operators 
$J^i_{\pm \alpha}=\vv_i \cdot \sigmav \pm  \alpha_i$
with $\vv_i \in \real^3$, $\| \vv_i \|=1$, and $\sigmav$ the vector formed by the Pauli matrices 
$\sigma_x$, $\sigma_y$, and $\sigma_z$): then 
$\kappa_{\rm{QJ}} (\alpha)=2\sum_i \gamma_i \min \{ \alpha_i^2,1\}$ for real $\alpha_i$'s.
One reaches perfect entanglement protection ($\overline{C(t)}=C_0$) only for vanishing
laser intensities $\alpha_i^2$. 
In the case of two qubits coupled to two baths at positive temperatures, 
a general choice of jump operators such that the density matrix 
(\ref{eq-def_rho}) satisfies the master equation
(\ref{eq-Lindblad}) with the four Lindblad operators $\sigma_{\pm}^i$, $i=A,B$, is 
$J_{\mu,\pm \alpha}^i = J_\mu^i \pm \alpha_\mu^i$ with
the jump rates $\gamma_{\mu,\pm \alpha}^i= \gamma_\mu^i/2$,
laser amplitudes $\alpha_\mu^i \in \complex$, and $J_\mu^i=
\sum_{m=\pm} (\gamma^i_{m}/\gamma_\mu^i)^{\frac{1}{2}} u_{\mu m}^i \sigma_m^i$ for an arbitrary 
unitary matrix $(u_{\mu m}^i)_{\mu=1,\cdots, N}^{m=\pm}$ 
(see  the discussion in the preceding section). 
The corresponding disentanglement rate,
$\kappa_{\rm{QJ}} (\alpha)= \sum_{\mu,i} \gamma_\mu^i  
[ \tr_{\complex^2} (J_\mu^{i \dagger} J_\mu^i ) +  2 | \alpha_\mu^i |^2 
- 2 | \det_{\complex^2} (J_\mu^i )
 + (\alpha_\mu^i )^2 |]/2$, is  equal to  $\kappa_{\rm{QJ}} (0)$
if $\det ( J_\mu^i)=0$ or for complex laser amplitudes 
$\alpha_\mu^i = |\alpha_\mu^i| e^{\I \theta_\mu^i}$
satisfying $2\theta_\mu^i = \arg (\det ( J_\mu^i))$; otherwise, $\kappa_{\rm{QJ}} (\alpha)$ is larger then 
 $\kappa_{\rm{QJ}} (0)$. We can conclude that the smallest disentanglement rate 
is given by (\ref{eq-optimal_kappa_T>0}) and
the best unravelings  to protect
the entanglement of the qubits are either the QJ model with jump operators
$J_1^i \propto (\gamma_+^i)^{\frac{1}{2}} \sigma_+^i + (\gamma_-^i)^{\frac{1}{2}} \sigma_-^i$
and $J_2^i \propto (\gamma_+^i)^{\frac{1}{2}} \sigma_+^i - (\gamma_-^i)^{\frac{1}{2}} \sigma_-^i$
or the corresponding homodyne unraveling with laser phases 
$\theta_1^i=\pi/2$ and $\theta_2^i = 0$.

Let us now consider a general QJ model with jump operators
$J_m^i$. A new unraveling of (\ref{eq-Lindblad}) is obtained from the QJ 
model with jump operators 
$J_{m,\pm \alpha}^{i} = J_{m}^i \pm  \alpha^i_m$ and rates $\gamma_{m,\pm \alpha}^{i} 
= \gamma_m^i/2$.
For large positive laser amplitudes  $\alpha_m^i \gg 1$,
this  dynamics converges after an appropriate coarse graining in time
to the QSD model described by the stochastic Schr\"odinger equation~\cite{Wiseman93,moi+Orszag}
\begin{eqnarray} \label{eq-QSD}
\nn
& & 
\displaystyle \ket{\D \psi} 
 =  
  \Bigl[ ( -\I H_0 - K)\D t 
  +
  \sum_{m,i}  \Bigr( \sqrt{\gamma^i_m } \bigl( J^{i}_m -  \Re \langle J^{i}_m \rangle_{\psi} 
\bigr) 
\\
& & 
 \times \D w_{m}^i
+ \gamma^i_m  \Bigl( \Re \langle J_{m}^{i} \rangle_{\psi}\, J_{m}^{i}
- \frac{(\Re \langle J^{i}_m \rangle_{\psi})^2}{2}  \Bigr)\D t  
\Bigr)
\Bigr] \ket{\psi}
\end{eqnarray}  
where $\D w_{m}^i$  are the It\^o differentials for independent real Wiener processes
satisfying the It\^o rules $\D w_{m}^i \D w_{n}^j = \delta_{ij} \delta_{mn} \D t$.
One can determine the mean concurrence for the QSD model (\ref{eq-QSD}) by taking the limit
 of the mean concurrence for the QJ dynamics with jump operators $J_{m ,\pm \alpha}^i$. 
This gives again the 
exponential decay  (\ref{eq-main_result}) but with a new rate
\begin{equation} \label{eq-kappaQDS}
\kappa_{\rm{ho}} 
   =    
  \frac{ \tr_{\complex^4} ( K )}{2} - 
    \sum_{m,i}  \! \gamma_{m}^{i} 
    \Bigl(
    \Re  {\det}_{\complex^2} (J_m^i)  
     + \frac{1}{2} \bigl( \Im  {\trace}_{\complex^2} (J_m^i) \bigr)^2  
    \Bigr).
\end{equation}
In fact, if $2 \theta_{m,\pm \alpha}^i$ is the argument
of $\det ( J_{m}^i\pm \alpha_m^i ) = (\alpha_m^i)^2\pm \alpha_m^i \tr (J_m^i) + 
{\mathcal{O}} (1)$ then for $\alpha_m^i \gg 1$,  $\alpha_m^i >0$, one has
$e^{2 \I \theta_{m,\pm \alpha}^i} \sim  1 \pm \I \Im  \tr ( J_m^i ) /\alpha_m^i$. Using
(\ref{eq-kappa_m}), a short calculation gives  (\ref{eq-kappaQDS}).

Unlike $\kappa_{\rm{QJ}}$, $\kappa_{\rm{ho}}$  changes when the
operators $J_m^i$ in (\ref{eq-QSD}) acquire a phase factor, 
$J_m^i \rightarrow \E^{-\I \theta_m^i}J_m^i$. 
This arises for homodyne 
detection with complex laser amplitudes 
$\alpha_m^i=|\alpha_m^i| \E^{\I \theta_m^i}$, $|\alpha_m^i|\gg 1$. 
Minimizing over the laser phases $\theta_m^i$ yields 
\begin{eqnarray} \label{eq-kappa_opt_hom}
\nonumber
\kappa_{\rm{ho}}^{\rm{opt}}
& = & 
\frac{1}{2} \tr_{\complex^4} (K) - \sum_{m,i} \gamma_m^i 
\Bigl( \bigl|\det_{\complex^2} (J_m^i) - \frac{1}{4} \bigl( \tr_{\complex^2} 
(J_m^i) \bigr)^2 \bigr|
\\
& & 
+\frac{1}{4} \bigl| \tr_{\complex^2} (J_m^i)  \bigr|^2 
\Bigr)\,.
\end{eqnarray}
It is easy to show that $\kappa_{\rm{ho}}^{\rm{opt}} \leq \kappa_{\rm{QJ}}$, this inequality
being strict excepted if the two eigenvalues of $J_m^i$ have the same modulus
for all $(m,i)$.
Thus optimal
homodyne detection protects entanglement better than - or, if the aforementioned condition 
is fulfilled,
as well as  - photon counting.
Let us stress that the optimal measurements (in particular, the laser phases $\theta_m^i$
minimizing the rate $\kappa_{\rm{ho}}$) only depend on the Lindblad operators $J_m^i$ in the 
master equation (\ref{eq-Lindblad}) and are thus the same for all  initial states of the qubits.

Let us now discuss the case of heterodyne detection. 
The corresponding jump operators 
$J^i_{m,\pm \alpha} (t_q) = J^i_m \pm \alpha_m^i e^{\I \Omega_m^i t_q }$ depend on
the time $t_q$  of the $q$-th jump due to the 
oscillations of the laser amplitudes~\cite{Knight98}. The associated rates are 
$\gamma_{m,\pm \alpha}^i = \gamma_m^i /2$ 
as for homodyne detection. We assume here that $\alpha_m^i >0$.
In the  limit $(\alpha_m^i)^2 \gg \Omega_m^i/\gamma_m^i \gg 1$  
of large laser intensities and rapidly oscillating laser amplitudes,
the QJ dynamics with jump operators $J^i_{m,\pm \alpha} (t_q)$ converges 
to the QSD model given by the stochastic Schr\"odinger equation~\cite{Breuer}   
\begin{eqnarray} \label{eq-heter}
\nn
& & 
\displaystyle \ket{\D \psi} 
 =  
  \Bigl[( -\I H_0 - K)\D t 
  +
 \frac{1}{2} \sum_{m,i} 
\gamma_m^{i} \Bigl( \langle J_{m}^{i} \rangle_{\psi}^\ast \, J_{m}^{i}
\\
\nn
& &
- \frac{1}{2} \bigl| \langle J^{i}_m \rangle_{\psi} \bigr|^2 \Bigr)\D t 
+ \sum_{m,i} \sqrt{\gamma^i_m } 
\Bigl( 
\bigl( J^{i}_m -  \frac{1}{2} \langle J^{i}_m \rangle_{\psi} \bigr) \D \xi_m^i 
\\
& & 
- \frac{1}{2} \langle J^{i}_m \rangle_{\psi}^\ast  ( \D \xi_m^i)^\ast   
\Bigr)
\Bigr] \ket{\psi}
\end{eqnarray}  
where $\D \xi_m^i$ are the It\^o differential of independent 
complex Wiener processes satisfying the It\^o rules $\D \xi_m^i \D \xi_n^j =0$ and
$\D \xi_m^i ( \D \xi_n^j)^\ast =\delta_{ij} \delta_{mn} \D t$.
Eq. (\ref{eq-heter}) describes the coarse-grained evolution of the normalized wavefunction
$\ket{\psi (t)}$ 
on a time scale $\Delta t$ such that (i) many jumps and many laser amplitude oscillations
occur in a time interval of length 
$\Delta t$ and (ii) $\ket{\psi (t)}$ does not change significantly on such a time interval.
These conditions are satisfied when 
$(\alpha_m^i )^2 \gamma_m^i \Delta t \gg  \Omega_m^i \Delta t  \gg 1$
and $\gamma_m^i \Delta t \ll 1$. 
We now show that
the mean concurrence  for the QSD model
(\ref{eq-heter}) is given by (\ref{eq-main_result}) 
and determine the rate $\kappa$ of its exponential decay.
This can be done by calculating the derivative
$\D \overline{C (t)}/\D t$ in a similar way as in 
Sec.\ \ref{sec-proof_main_result}, using (\ref{eq-heter}) and the It\^o rules.
It turns out to be simpler to estimate directly the average concurrence of the 
QJ model for heterodyne detection in the aforementioned limits, in analogy with our
previous analysis for homodyne detection. 
 Let us first remark that the results of 
Sec.\ \ref{sec-proof_main_result} remain valid
if the jump operators $J_m^i(t)$ vary slowly in time, on a time scale 
$(\Omega_m^i)^{-1}$ much larger than the mean time $(\alpha_m^i )^{-2}/\gamma_m^i$ 
between consecutive jumps. 
Hence 
$\D \overline{C} /\D t = - \kappa_{\rm{het}} (t)\, \overline{C}(t)$ and thus
$\overline{C(t)} = C_0 \, e^{- \int_0^t \D t'\, \kappa_{\rm{het}} (t')}$
with a time-dependent rate $\kappa_{\rm{het}} (t)$ 
given by (\ref{eq-kappa_m}). To simplify notations, we temporarily  omit 
the sum in  (\ref{eq-kappa_m}) and do not write explicitly
 the lower and upper indices $m$ and $i$. Let us set
$\tau = \tr ( J )/2 = |\tau| e^{\I \varphi}$ and
$\delta = \det (J) = e^{2\I \theta} | \delta|$. Let 
   $2 \theta_{\pm \alpha} (t)$ denote the argument of $\det ( J \pm \alpha e^{\I \Omega t})$. 
Generalizing the calculation outlined above for homodyne detection, one gets 
$e^{2 \I \theta_{\pm \alpha} (t)} \sim 
e^{2 \I \Omega t} ( 1 \pm 2 i \Im \{ \tau\, e^{-\I \Omega t} \}/\alpha)$
as $\alpha \gg 1$. By (\ref{eq-kappa_m}) this yields  
\begin{eqnarray*}
& & \kappa_{\rm{het}} (t)
 =  
\frac{\gamma}{2} 
 \Bigl( 
  \bigl| 
   \braup {J} \ketup - e^{2 \I \Omega t} \bradown J^{\dagger} \ketdown
\\
& & 
\hspace*{1.2cm} - 2 \I e^{\I \Omega t}\, \Im \{ \tau e^{-\I \Omega t} \}  
\bigr|^2 + 
\bigl| 
 \braup \bigl( {J} +  e^{2 \I \Omega t } J^{\dagger} \bigr) \ketdown 
\bigr|^2
\Bigr)
\\
& & = 
\frac{\tr_{\complex^4} (K)}{2} - \gamma
 \bigl(
    | \delta | \cos ( 2 \theta - 2 \Omega t ) + 2 | \tau |^2 \sin^2 (\varphi - \Omega t ) 
\bigr)
 \end{eqnarray*}
up to terms of order $\alpha^{-1}$.
By neglecting
the oscillatory integral  
$\int_{0}^{t} \D t'\,\cos (2  \theta - 2 \Omega t')$ 
(which is of order $\Omega^{-1} \ll \Delta t \leq t$) and approximating 
$\int_{0}^{t} \D t'\,\sin^2 ( \varphi - \Omega t')$ by $t/2$,
one obtains
$
\int_{0}^{t} \D t' \, \kappa_{\rm{het}} (t') \simeq
t 
(\tr_{\complex^4} ( K) /2 - \gamma  |\tau |^2  )
$.
Putting together the previous results, 
this shows that  $\overline{C(t)} \rightarrow C_0 e^{-\kappa_{\rm{het}} t}$  
in the limit $\alpha^2 \gg \Omega/\gamma \gg 1$ and
$\Omega \gg (\Delta t)^{-1} \gg \gamma$, with 
\begin{equation} \label{eq-kappa_het}
\kappa_{\rm{het}} 
   =    
  \frac{ \tr_{\complex^4} ( K )}{2} - 
   \frac{1}{4}  \sum_{m,i}   \gamma_{m}^{i} 
    \bigl| \tr(J_m^i) \bigr|^2
    \, .
\end{equation}
We note that $\kappa_{\rm{het}}  \geq \kappa_{\rm{ho}}^{\rm{opt}}$. 
For given jump operators $J_m^i$, the measurement scheme which better protects
the qubits against disentanglement is thus given by homodyne detections with 
optimally chosen laser phases. In this scheme, the average concurrence decays 
exponentially 
with the rate (\ref{eq-kappa_opt_hom}).

Although  (\ref{eq-heter}) is different from the QSD equation for the 
normalized wavefunction introduced 
by Gisin and Percival~\cite{Gisin92}, the quantum trajectories $t \mapsto \ket{\psi (t)}$
for the two dynamics are the same  
up to a random fluctuating phase~\cite{Breuer} which does not affect the concurrence
$C_{\psi(t)}$. More generally, one can show that the mean concurrence for the QSD model
with correlated complex noises satisfying the It\^o rules
 $\D \xi_m^i \D \xi_n^j =u_{mn}^{ij} \D t$ and
$\D \xi_m^i ( \D \xi_n^j)^\ast =\delta_{ij} \delta_{mn} \D t$~\cite{Wiseman01}, which gives
back the model of Gisin and Percival when $u_{mn}^{ij}=0$, decays exponentially as
in (\ref{eq-main_result}) if the two baths are independent, \ie, if $u_{mn}^{AB}=0$ for any 
$m,n$.

%%%%%%%%%%%%%%%FIGURE 2%%%%%%%%%%%%%%%%%%%%%%%%%%%%%%%%%%%%%%%%%%
\begin{figure}
\includegraphics*[width=0.99\columnwidth]{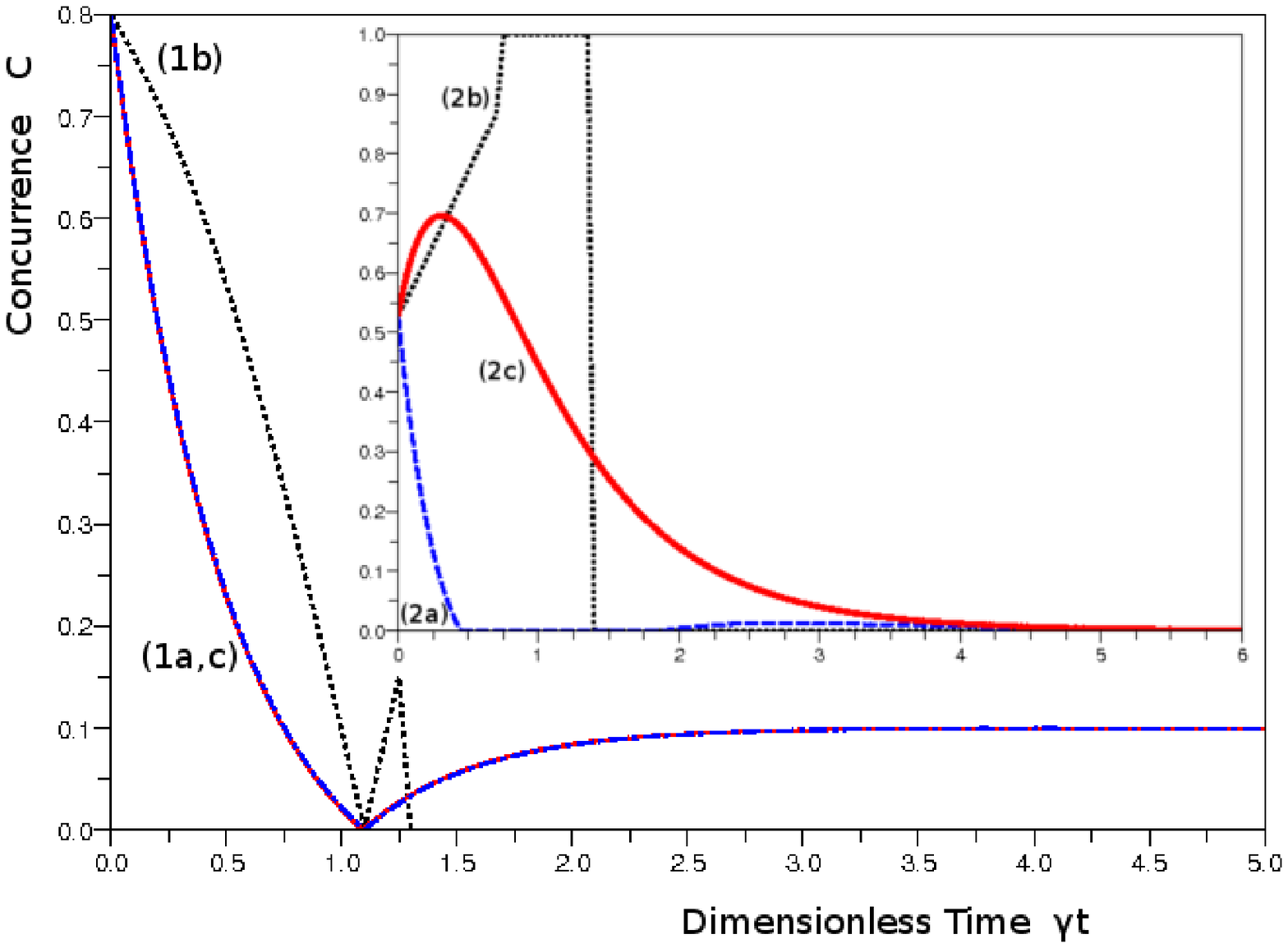}
\caption{(Color online) \label{fig_common_bath}
Concurrence of two qubits coupled to a common bath versus $\gamma t$ for 
$\ket{\psi(0)}=\frac{2}{\sqrt{5}} \ketupdown + \frac{1}{\sqrt{5}} \ketdownup$:
(1a)~$C_{\rho(t)}$  (blue dashed line);
(1b)~$C_{\psi (t)}$ for a single trajectory (black dotted line);
(1c)~$\overline{C_{\psi (t)}}$ given by~(\ref{eq-result_1_bath}) 
(red line superimposed on the blue line).
Inset (2) is the same for 
$\ket{\psi(0)}=\frac{7\I}{\sqrt{53}} \ketupup + \frac{2\I}{\sqrt{53}} 
\ketdowndown$. }
\end{figure}
%%%%%%%%%%%%%%%%%%%%%%%%%%%%%%%%%%%%%%%%%%%%%%%%%%%%%%%%%%%%

%%%%%%%%%%%%%%%%%%%%%%%%%%%%%%%%%%%%%%%%%%%%%%%%%%%%%%%%%%
\section{Qubits coupled to a common bath}
\label{sec-common_bath}
%%%%%%%%%%%%%%%%%%%%%%%%%%%%%%%%%%%%%%%%%%%%%%%%%%%%%%%%%%%%

We focus here on a specific model of two qubits with equal frequencies
coupled resonantly to
the same modes of the electromagnetic field initially in the vacuum.
A photon counter $D$ makes a click when a photon is emitted by qubit $A$ or $B$. 
The  jump operator  in the 
rotating wave approximation, $J= \sigma_{-} \otimes 1_B+1_A \otimes \sigma_{-}$,
is now non-local. We take $H_0=0$.
Proceeding as for independent baths, 
the contribution to the mean concurrence  of quantum trajectories without jump 
 between  $0$ and $t$ is $p_{\rm nj} (0,t) C_{\rm nj} (t)=
|\langle \sigma_y \otimes \sigma_y T \rangle_{\E^{-t K} \ket{\psi(0)}}|$ and can be 
determined with the help of (\ref{eq-D(t)}).
By calculating the exponential of $K=\gamma J^\dagger J/2$, one finds  
$e^{- (t-t_0) K} \ket{\psi (t_0)} = \sum_{s,s'} c_{s s'} (t) \ket{s,s'}$ with
$c_{\up \up} (t) = e^{-\gamma (t-t_0)} c_{\up \up} (t_0)$,
$2 c_{s s'} (t) = (e^{-\gamma (t-t_0)} + 1 ) c_{s s'} (t_0) 
+ (e^{-\gamma (t-t_0)} - 1 ) c_{s's } (t_0)$ for $s s'=\up \down$ or $\down \up$, 
and $c_{\down \down}(t) = c_{\down \down}(t_0)$.  
Quantum trajectories  having one jump in $[0,t]$ give a nonzero contribution.
The probability density 
 that the jump occurs at time $t_j \in [0,t]$ is given by 
$\gamma p_{\rm{nj}} (t_j,t) \| J \ket{\psi (t_j-)}\|^2  p_{\rm{nj}} (0,t_j) 
 =\gamma\, \Nn_{\text{1j}, t_j} (t)^{2}$ with
$\Nn_{\text{1j}, t_j} (t) =\| \E^{-(t-t_j)K} J \E^{-t_j K} \ket{\psi (0)}\|$
(this follows from the formula
$p_{\rm{nj}} (t_0,t)= \| e^{-(t-t_0) K} \ket{\psi (t_0)}\|^2$, see Sec.\ \ref{sec-QJM}).
The contribution of trajectories  having one jump in $[0,t]$ 
 is then obtained by multiplying this density by
$C_{\text{1j}, t_j} (t) = 2 \Nn_{\text{1j}, t_j} (t)^{-2} \E^{-2 \gamma t} | c_{\up \up}|^2$
and integrating over $t_j$. 
After two clicks, $\ket{\psi (t)} = \ketdowndown$ is in an invariant separable state.
Therefore, trajectories with more than one jump do not contribute to the mean concurrence. 
Setting $c_\pm = c_{\up \down}\pm c_{\down \up}$, one gets
\begin{equation} \label{eq-result_1_bath}
\overline{C(t)} \!=\!  
\frac{1}{2}\bigl| c_{-}^2
\!-  c_{+}^2 e^{-2 \gamma t}
\!+  4 c_{\uparrow \uparrow} c_{\downarrow \downarrow}e^{-\gamma t}
\bigr| +2| c_{\uparrow \uparrow}|^2 \gamma te^{-2\gamma t} .
\end{equation}
The time behavior of the concurrence (\ref{eq-result_1_bath})
depends strongly on the initial state. 
Unlike in the case of independent baths, $\overline{C(t)}$
may vanish at nonzero finite discrete times $t_0$. 
A necessary and sufficient condition for this 
loss of entanglement (immediately followed by a revival) is  
$c_{\up \up}=0$ and $\arg (c_{\up\down}) = \arg (c_{\down \up}) $ 
(\ie, $c_{+}/c_{-} \in ]-\infty,-1[ \cup ]1,\infty[$).
If this condition is fulfilled,
$\overline{C(t)}$
 vanishes at time $t_0=\gamma^{-1} \ln ( |c_+/c_-|)$, see 
Fig.~\ref{fig_common_bath}. It is not difficult to show by solving the master equation
(\ref{eq-Lindblad}) with $J= \sigma_{-} \otimes 1_B+1_A \otimes \sigma_{-}$
that, for any initial state containing at most one excitation (\ie, such that
$c_{\up \up}=0$), 
$\overline{C(t)}=| c_-^2 - c_+^2 e^{-2 \gamma t} |/2$ coincides at all times with
the concurrence $C_{\rho(t)}$ for the density matrix.
In contrast,  if $c_{\up \up}\not = 0$ then  
$\overline{C(t)}$ increases at small times 
whereas $C_{\rho(t)}$ decreases,
as shown in the inset of Fig.~\ref{fig_common_bath}. 
For any 
initial state, $\overline{C(t)}$ converges at
large times $t \gg \gamma^{-1}$ to the same asymptotic value
$C_\infty = | c_-|^2/2$ as the  concurrence $C_{\rho(t)}$~\cite{Maniscalco08,Orszag10}. 

A non-local measurement scheme depending on the initial state $\ket{\psi (0)}$ 
and such that 
$\overline{C (t)} = C_{\rho(t)}$ at all times $t \in [0,t_{\rm EDS}]$
has been found recently~\cite{Viviescas-arXiv} 
for two qubits coupled to two baths at zero temperature in the rotating-wave approximation.
If one neglects the Hamiltonian of the qubits, this  scheme 
is time-independent. The corresponding quantum trajectories are given by a 
QSD equation~\cite{Wiseman01} for homodyne detection
with  two jump operators $J_1$ and $J_2$ similar to the jump operator $J$ 
introduced in this section,
combined with intense laser fields via 50\% beam splitters,
as described in Sec.\ \ref{sec-QSD}   
(the main difference between $J_{1,2}$ and $J$ comes from the presence of 
appropriately chosen phase factors in front of $\sigma_{-}$ and $\sigma_+$
making $J_{1,2}$  non-symmetric under the exchange of 
the two qubits). It is striking that       
we also find  in our model that $\overline{C (t)} = C_{\rho(t)}$ 
for specific initial states even though the  dynamics 
 in the absence of measurements - and thus the density matrix concurrence
 $C_{\rho(t)}$ -  are not the same in the two models  
(here the two qubits are coupled to a common bath,
whereas they are coupled to distinct baths in Ref.~\cite{Viviescas-arXiv}).

%%%%%%%%%%%%%%%%%%%%%%%%%%%%%%%%%%%%%%%%%%%%%%%%%%%%%%%%%%%%%%
\section{Conclusion}
\label{sec-conclusion}
%%%%%%%%%%%%%%%%%%%%%%%%%%%%%%%%%%%%%%%%%%%%%%%%%%%%%%%%%%%%%%%%%

 We have found explicit formulas for the mean concurrence
$\overline{C(t)}$ of quantum trajectories and have 
shown that the measurements on the baths may be used to protect
the entanglement of two qubits. 
These results shed new light on the phenomenon of entanglement sudden death. 
For independent baths, $\overline{C(t)}$ is either constant in time or 
vanishes exponentially
with a rate depending on the measurement scheme only, whereas for 
a common bath $\overline{C(t)}$ depends strongly on the
initial state and may coincide with the concurrence $C_{\rho (t)}$ of 
the density matrix for some initial states.
A constant $\overline{C(t)}$ implies a perfect protection of
maximally entangled states for all quantum trajectories.
In the case of pure dephasing and for Jaynes-Cumming couplings at infinite temperature,
we have found measurement schemes independent of the  initial state of the qubits which
lead to such a perfect entanglement protection.   
Despite obvious analogies, this way to protect entanglement 
differs from the strategy based on the quantum Zeno effect proposed 
in Ref.~\cite{Maniscalco08}. 
In fact, in the QJ and QSD models considered 
here the time interval  between consecutive measurements
is not arbitrarily small with respect to the damping constant 
$\gamma^{-1}$. In the QJ model  
this time interval 
$\D t$ must be chosen 
such that the jump probability $\D p (t) \propto \gamma\, \D t$ 
is very small but one cannot let $\gamma \, \D t$ go to zero since this would amount
to replacing $\D p (t)$ by $0$ and $e^{-\I H_{\rm{eff}} \D t}$ by $e^{-\I H_0 \D t}$ in   
(\ref{eq-no_jump}). In contrast, a perfect entanglement protection 
is reached in~\cite{Maniscalco08}  in the idealized 
limit $\gamma\, \D t \rightarrow 0$ (\ie, when  the
measurements completely prevent the decay of the superradiant state~\cite{Fisher01}).

For independent baths, $\overline{C(t)}$ is strictly greater 
than $C_{\rho (t)}$
if the latter concurrence vanishes after a finite time. Therefore, if there exists 
a measurement scheme such that the mean entanglement of formation
$\overline{E (t)}$ is equal to the entanglement of formation of the density
matrix (which would imply $\overline{C(t)} \leq C_{\rho (t)}$),
this scheme must necessarily involve measurements of 
non-local (joint) observables of the two baths. 
Let us finally note that it should be possible to check our findings 
experimentally by using similar optical devices as in Ref.~\cite{Almeida07}.   

\vspace{2mm}

%=====================================================================================
\begin{center}
{\bf{ACKNOWLEDGMENTS}}
\end{center}
We thank  P. Degiovanni, A.~Joye, and C. Viviescas for interesting discussions. 
We acknowledge financial support from the Agence Nationale de la Recherche 
(Grant No. ANR-09-BLAN-0098-01).
 
\vspace{3mm}

\noindent {\bf Note added:} after the completion of this work we learned
that related results have been obtained in~\cite{Santos-arXiv}.

%=====================================================================================


\begin{thebibliography}{33}



\bibitem{Diosi04}  L. Di\'osi, in {\it Irreversible Quantum Dynamics},
{{Lecture Notes in Physics} {\bf 622}, 157, edited by F. Benatti and R. Floreanini (Springer, Berlin, 2003)

\bibitem{Dodd04}
P.J. Dodd and J.J. Halliwell, Phys. Rev. A {\bf 69}, 052105 (2004)


\bibitem{Eberly04} T. Yu and J.H. Eberly,  Phys. Rev. Lett. 
{\bf  93}, 140404 (2004)


\bibitem{Almeida07} M.P. Almeida {\it et al.}, Science {\bf 316}, 579 (2007)



\bibitem{Terra_Cunha07} M.O. Terra Cunha, New J. Phys. {\bf 9}, 237 (2007)

\bibitem{Braun02} D. Braun, {Phys. Rev. Lett.} {\bf 89}, 277901 (2002) 

\bibitem{Ficek06} Z. Ficek and R. Tan\'as}, {Phys. Rev. A} {\bf 74}, 024304 (2006)

\bibitem{Mazzola09} L. Mazzola, S. Maniscalco, J. Piilo, K.-A. Suominen, and B.M. Garraway,
{Phys. Rev. A} {\bf 79}, 042302 (2009)


\bibitem{Dalibard92}  J. Dalibard,  Y. Castin,  and K. M{\o }lmer, Phys. Rev. Lett. 
\textbf{68}, 580 (1992) 

\bibitem{Carmichael} H.J. Carmichael, \emph{An Open System Approach to
Quantum Optics}, Lecture Notes in Physics, Series {\bf{M18}} (Springer, Berlin, 1993)
 


\bibitem{Wiseman93} H.M. Wiseman and G.J. Milburn, {Phys. Rev. A} {\bf 47}, 642 (1993)


\bibitem{Gisin92}   N. Gisin and  I.C. Percival, J. Phys A: Math. Gen.
\textbf{25}, 5677 (1992)

\bibitem{Nha04} H. Nha and H.J. Carmichael, {Phys. Rev. Lett.} {\bf 93}, 120408 (2004)

\bibitem{Carvalho}
A.R.R. Carvalho, M. Busse, O. Brodier, C. Viviescas, and A. Buchleitner, {Phys. Rev. Lett.}
{\bf 98}, 190501 (2007)

\bibitem{Maniscalco08} S. Maniscalco, F. Francica, R.L. Zaffino, N. Lo Gullo, 
and F. Plastina, Phys. Rev. Lett. {\bf 100}, 090503 (2008)

\bibitem{Mundarain09} D.M. Mundarain and M. Orszag, {Phys. Rev. A} {\bf 79}, 052333 (2009)

\bibitem{CarvalhoPRA07} A.R.R. Carvalho and J.J. Hope, {Phys. Rev. A} {\bf 76}, 010301(R) (2007)

\bibitem{Mascarenhas10} E. Mascarenhas,
B. Marques, D. Cavalcanti, M.O. Terra Cunha, and M. Fran\c{c}a Santos, 
Phys. Rev. A {\bf 81}, 032310 (2010) 

\bibitem{Viviescas-arXiv} C. Viviescas, I. Guevara, A.R.R. Carvalho, M. Busse, 
and A. Buchleitner, arXiv:1006.1452v1 [quant-ph]

\bibitem{Bennett96} 
C.H. Bennett, D.P. DiVincenzo, J.A. Smolin, and W.K. Wootters,
{Phys. Rev. A} {\bf 54}, 3824 (1996)


\bibitem{Wootters98} W.K. Wootters, {Phys. Rev. Lett.} {\bf 80}, 2245 (1998) 



\bibitem{Breuer} H.-P. Breuer and F. Petruccione, {\it The Theory of Open Quantum Systems}
(Oxford University Press, 2002) 

\bibitem{Knight98} M.B. Plenio and P.L. Knight, {Rev. Mod. Phys.} 
{\bf{70}}, 101  (1998) 
  

\bibitem{Haroche}  S. Haroche and J.-M. Raimond, {\it Exploring the quantum: atoms, 
cavities and photons} (Oxford Univ. Press, 2006)


\bibitem{moi+Orszag} D. Spehner and M. Orszag,   J. Math. Phys. {\bf 43}, 3511 (2002) 

\bibitem{Wiseman01} H.M. Wiseman and L. Di\'osi, Chem. Phys. {\bf 268}, 91 (2001)

\bibitem{Orszag10} M. Orszag and M. Hernandez, Adv. in Optics and Photonics {\bf 2}, 229 (2010)

\bibitem{Fisher01} M.C. Fischer, B. Gutierrez-Medina, and
  M.G. Raizen, {Phys. Rev. Lett} {\bf 87}, 040402  (2001); 
 P.E. Toscheck and C.~Wunderlich, {Eur. Phys. J.}
  D {\bf 14}, 387  (2001)

\bibitem{Santos-arXiv} E.~Mascarenhas, D.~Cavalcanti, 
V.~Vedral, and M.~Fran\c{c}a Santos, arXiv:1006.1233 [quant-ph]


\end{thebibliography}
\end{document}